\documentclass[twocolumn]{aastex63}

\newcommand{\eg}{e.\,g.,}
\newcommand{\dmunits}{pc~cm$^{-3}$}
\usepackage{color}
\usepackage{graphicx, epstopdf} 
\usepackage{gensymb}
\usepackage{amsmath}
\usepackage{float}

\newcommand{\rf}{{\sc Realfast}}

\graphicspath{{./}{figures/}}

\received{January 1, 2018}
\revised{January 7, 2018}
\accepted{\today}
\submitjournal{ApJ}

\shorttitle{Clustering Analysis}
\shortauthors{Aggarwal et al.}

\begin{document}

\title{Robust Assessment of Clustering Methods for Fast Radio Transient Candidates}

\correspondingauthor{Kshitij Aggarwal}
\email{ka0064@mix.wvu.edu}

\author[0000-0002-2059-0525]{Kshitij Aggarwal}
\affil{West Virginia University, Department of Physics and Astronomy, P. O. Box 6315, Morgantown, WV, USA}
\affil{Center for Gravitational Waves and Cosmology, West Virginia University, Chestnut Ridge Research Building, Morgantown, WV, USA}

\author[0000-0003-4052-7838]{Sarah Burke-Spolaor}
\affil{West Virginia University, Department of Physics and Astronomy, P. O. Box 6315, Morgantown, WV, USA}
\affil{Center for Gravitational Waves and Cosmology, West Virginia University, Chestnut Ridge Research Building, Morgantown, WV, USA}
\affiliation{CIFAR Azrieli Global Scholars program, CIFAR, Toronto, Canada}

\author[0000-0002-4119-9963]{Casey J. Law}
\affiliation{Cahill Center for Astronomy and Astrophysics, MC 249-17 California Institute of Technology, Pasadena, CA 91125, USA}

\author[0000-0003-4056-9982]{Geoffrey C. Bower}
\affiliation{Academia Sinica Institute of Astronomy and Astrophysics, 645 N. A'ohoku Place, Hilo, HI 96720, USA}

\author[0000-0002-5344-820X]{Bryan J. Butler}
\affiliation{National Radio Astronomy Observatory, Socorro, NM, 87801, USA}

\author[0000-0002-6664-965X]{Paul B. Demorest}
\affiliation{National Radio Astronomy Observatory, Socorro, NM, 87801, USA}

\author{T. Joseph W. Lazio}
\affiliation{Jet Propulsion Laboratory, California Institute of Technology, 4800 Oak Grove Dr, M/S 67-201, Pasadena, CA 91109 USA}

\author[0000-0002-3873-5497]{Justin Linford}
\affiliation{National Radio Astronomy Observatory, Socorro, NM, 87801, USA}

\author[0000-0002-3360-9299]{Jessica Sydnor}
\affil{West Virginia University, Department of Physics and Astronomy, P. O. Box 6315, Morgantown, WV, USA}
\affil{Center for Gravitational Waves and Cosmology, West Virginia University, Chestnut Ridge Research Building, Morgantown, WV, USA}

\author[0000-0001-8057-0633]{Reshma Anna Thomas}
\affil{West Virginia University, Department of Physics and Astronomy, P. O. Box 6315, Morgantown, WV, USA}
\affil{Center for Gravitational Waves and Cosmology, West Virginia University, Chestnut Ridge Research Building, Morgantown, WV, USA}

\begin{abstract}
Fast radio transient search algorithms identify signals of interest by iterating and applying a threshold on a set of matched filters. These filters are defined by properties of the transient such as time and dispersion. 
A real transient can trigger hundreds of search trials, each of which has to be post-processed for visualization and classification tasks. In this paper, we have explored a range of unsupervised clustering algorithms to cluster these redundant candidate detections. 
We demonstrate this for \rf, the commensal fast transient search system at the Very Large Array. We use four features for clustering: sky position (\emph{l}, \emph{m}), time and dispersion measure (DM). We develop a custom performance metric that makes sure that the candidates are clustered into a \emph{small} number of \emph{pure} clusters, i.e, clusters with either astrophysical or noise candidates. We then use this performance metric to compare eight different clustering algorithms. 
We show that using sky location along with DM/time improves clustering performance by $\sim$10\% as compared to the traditional DM/time-based clustering. Therefore, positional information should be used during clustering if it can be made available. 
We conduct several tests to compare the performance and generalisability of clustering algorithms to other transient datasets and propose a strategy that can be used to choose an algorithm. Our performance metric and clustering strategy can be easily extended to different single-pulse search pipelines and other astronomy and non-astronomy-based applications. 

\end{abstract}

\keywords{Clustering(1908) --- Random Forests(1935) --- Radio transient sources(2008) ---
Radio interferometry(1346) --- Extragalactic radio sources(508) --- Radio bursts(1339) --- Very Large Array(1766)}

\section{Introduction}
\label{sec:intro}





One of the significant difficulties when seeking fast-transient radio signals is the load of candidates that results from a transient search: it is common for a search algorithm to return millions to billions of candidates from a survey, only a few of which end up being genuine (the rest being thermal noise and radio-frequency interference---RFI). Even one bright event
, whether astrophysical or artificial, can generate many hundreds of separate candidates. This is because search algorithms iterate over a set of matched filters and identify transients that exceed the detection threshold. Clustering algorithms to account for this effect are of dire importance to any radio transient search pipeline. A rigorous study of an effective clustering algorithm for fast-radio-transient searches is the primary purpose of the study reported here.

%

To understand this paper's context, it is important to review the main procedural components of a typical search for fast radio transients. The term ``fast'' here specifically refers to transients for which the dispersion delay, caused by astrophysical plasma, is non-negligible and must be accounted for to optimize search sensitivity. 
The tenuous plasma that fills the space between stars, around galaxies, between galaxies, and elsewhere can have a strong influence on radio signals. The most prominent influence they have is inducing a frequency-dependent pulse sweep caused by a frequency-dependent refractive index of cold astrophysical plasma \citep{handbook}. The magnitude of this dispersive time delay for a pulse is quantified by the \textit{dispersion measure} (DM).


When searching for a radio transient, neither its dispersion measure nor width (i.e., duration) are known.
Therefore, one must search over a range of DMs and widths to carry out a full-sensitivity search. DM values at which to search are chosen by considering the expected decline in signal-to-noise (S/N) ratio, due to pulse broadening, at the adjacent DMs \citep{cordes2003, levin2012}. To summarise, a standard fast-transient-search pipeline dedisperses the data at various trial DMs, averaging all the frequencies to obtain a one-dimensional time series, followed by convolution using boxcar filters of various widths.
Candidate pulses are identified by searching for peaks above a pre-decided threshold, with the S/N of a candidate determined from an estimate of the signal strength with respect to the standard deviation within the region defined by the boxcar filter width.  
Dedispersing at an incorrect DM, or using a boxcar filter of incorrect width, would reduce the S/N of the pulse. As previously noted, any event can lead to multiple candidates being detected by the search pipeline if the S/N remains above the threshold at the incorrect DM or boxcar width.

This process can lead to a substantial number of redundant candidates caused by a single event. Clustering is performed on these candidates to automatically combine such events at the end of the search pipeline. 
Some algorithms that are currently in common use are friends of friends (FoF) and DBSCAN \citep[][]{dbscan,Deneva2009,barsdell2012}. However, few clustering algorithms have been rigorously tested.


Throughout this work, our primary motivation was to identify the optimal clustering algorithm for single-pulse searches, in particular in searches for fast radio bursts (FRBs). FRBs are bright, millisecond-duration bursts of energy of extragalactic origin \citep{lorimer07}. Over 150 such sources have been seen so far \citep{petroff16}, and many radio telescopes worldwide (both single-dish and interferometric) are now or will soon be outfitted with specialized hardware and software to carry out FRB searches.


The \rf\ system at Very Large Array (VLA) is one such commensal fast transient search system \citep{law2018}. It is currently the only real-time coherent-imaging interferometric search system, although because of the importance of precise FRB localization, a number of similar systems are in operation, commissioning, or planning phases \citep{Kocz2019, bannister2019, Michilli2020, Leung2021}. \rf\ forms thousands of de-dispersed images every second to search for pulses in the image plane and can localize every FRB it detects to arcsecond precision \citep[][]{law2018}. 

The prototype \rf\ system was used for the first localization of an FRB \citep{chatterjee17}. In its first year of commensal observation at L (1-2\,GHz), S (2-4\,GHz) and C-band (4-8\,GHz), \rf\ detected 5 FRBs \citep[Aggarwal et al. in prep; Bhandari et al. in prep; Tendulkar et al. in prep;][]{law2020, Aggarwal2020}. The \rf\ pipeline focuses on searching for transients at multiple DMs and trial widths, each of which is then post-processed. A total intensity (Stokes~I) image is formed for each trial DM and width. Point sources in these images with S/N greater than a pre-set threshold trigger the detection pipeline. The data corresponding to each candidate are then saved to disk and is classified using a Deep Learning based classifier \citep[][]{Agarwal2020}. Visualizations that show the radio image, spectrogram, spectra, and profile of the candidate are then generated. These visualizations also consist of other relevant candidate parameters: signal to noise ratio (S/N), \hbox{DM}, width, relative sky position with respect to the pointing center, scan number, etc., and are used for follow-up inspection.   

In this work, we use \rf\ data as a test-case to explore and compare candidate-clustering techniques. We also generalize our results to apply to single-dish telescopes, which do not have spatial (sky-location) information to use in a clustering algorithm. 
This paper is laid out as follows: In \S\ref{sec:clustering} we provide a more detailed motivation for the need for clustering and a discussion of clustering methodologies.
\S\ref{sec:data} describes the data used for testing the algorithm, followed by methods explained in \S\ref{sec:methods}. The results of the analysis are presented in \S\ref{sec:results}, followed by discussion and conclusion in \S\ref{sec:discussion} and \S\ref{sec:conclusions}, respectively.


\section{Clustering}\label{sec:clustering}

As mentioned previously, clustering is implemented between  the  search  and  the  candidate processing steps of the pipeline. In the \rf\ system, after clustering, we choose the maximum S/N candidate from each cluster, and only those are analyzed in the candidate processing step. We also consider all the unclustered candidates as individual clusters of size one and pass them onward for processing. 

In this section, we discuss the need to use clustering in the context of a single pulse search pipeline. Further, we use the following terminology\label{terminology} throughout this paper: 
\begin{itemize}
    \item Event: The actual physical occurrence of an astrophysical transient (\eg\ FRB, pulsar) or Radio Frequency Interference (RFI).
    \item Candidate: A single detection reported by a search pipeline. It typically consists of a set of properties (sky location, \hbox{DM}, time, etc). Candidates may be random thermal noise or associated with an event. Multiple candidates can be associated to a single event. 
    \item Observation: A set of candidates generated after the search pipeline is run on some data. It can be real or simulated and can have candidates associated with FRB or RFI or both.       
    \item Dataset: A set of observations.
    \item Cluster: A group of candidates (or members) with the same labels assigned by a clustering algorithm.
    \item Member: Candidates within a cluster.  
    \item True Labels: Each member of a cluster is associated with an event. We refer to this event as the true label of that member.
    \item Real/FRB/Transient: Event, cluster, or member associated with an astrophysical transient.
    \item RFI: Event, cluster, or member that is not astrophysical. 
\end{itemize}


\subsection{Expected number of candidates from a single astrophysical event}
\label{sec:whyclustering}

As explained in the previous section, following search parameters are reported for each candidate detected by the \rf\ search pipeline: \hbox{DM}, time of occurrence of the candidate, relative sky position with respect to the pointing center (\emph{l}, \emph{m}), \hbox{S/N}, and width.
Moreover, for each event, the pipeline can return multiple candidates at nearby (incorrect) values of \hbox{DM}, width, and time. 
The observed S/N of a candidate detected at a trial \hbox{DM}, width, sky location is given by (assuming no other losses, etc.)
\begin{equation}
    S/N_\text{observed} = S/N_\text{optimal}.F_\text{widthloss}.F_\text{beamloss}.F_\text{DMloss}    
\end{equation}

where $S/N_\text{optimal}$ is the S/N of the candidate when there is no loss. $F_\text{widthloss}$ is the loss due to incorrect boxcar width \citep{cordes2003}, $F_\text{beamloss}$ is the loss due to position of candidate within the primary beam of the telescope, and $F_\text{DMloss}$ is the loss due to incorrect DM value \citep[\S2.3]{levin2012}. 

\begin{figure}
    \centering
    \includegraphics[width=0.5\textwidth]{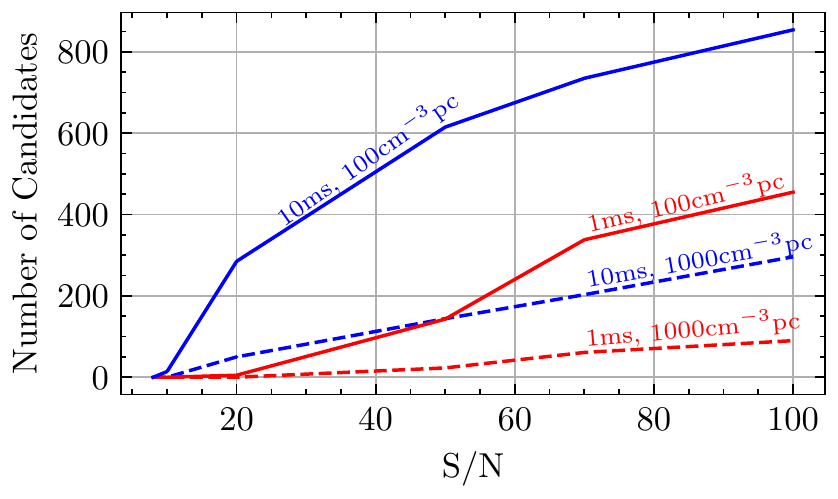}
    \caption{The number of candidates generated by a single pulse search pipeline for events with varying S/N. Different colors represent different intrinsic widths of the transient, and different line styles show different DMs. Observing and search configuration similar to that of \rf\ at L-band was chosen (Sampling time:\,5~ms, Number of frequency channels:\,256, Bandwidth:\,1~GHz, DM range:\,0 to 3,000\dmunits, Boxcar widths:\,[5~ms, 10~ms, 20~ms, 40~ms], S/N threshold:\,8). See Section~\ref{sec:whyclustering} for more details. } 
    \label{fig:numcands}
\end{figure}

Using this equation, we can calculate the number of candidates that will trigger a single pulse search system for an astrophysical event. For instance, we assume a VLA \mbox{L-band} (1--2~GHz) configuration with 256 frequency channels and a time resolution of 5~ms, and that \rf\ system is used to search for transients. We then search for DMs from~0~\dmunits\ to~3000~\dmunits\ and set $F_\text{dmloss} = 0.95$, i.e, up to a maximum of 5\% loss in sensitivity between DM trials. Using this, and assuming an intrinsic pulse width of 30~ms, we can compute the DM array \citep[\S2.3]{levin2012}.  We also assume $t_\text{scatt}=0$ as it is line-of-sight dependent and is typically small. We set our boxcar search widths to 5~ms, 10~ms, 20~ms, and~40~ms and S/N detection threshold at 8. Figure~\ref{fig:numcands} shows the number of candidates detected with respect to input S/N of the transient for two different values of transient DMs and widths. Here, we have also assumed that the candidate is at the center of the beam, therefore the $F_\text{beamloss}$ is 1.
This figure clearly shows that the number of candidates detected by the pipeline can be large even for one event. This can overwhelm the real-time systems that are responsible for post-processing these candidates and writing their data to disk, hence motivating the use of clustering algorithms.

\subsection{Unsupervised Clustering}
In this paper, we have taken the approach of unsupervised clustering. Here we briefly discuss unsupervised clustering and some of its caveats. Unsupervised clustering is the method used to find meaningful clusters from an unlabeled dataset, i.e., \textit{a priori} information about number of clusters and the true cluster assigned for each candidate is not known \citep[][]{jain1999}. This is in contrast to supervised clustering for which this information is available. In this analysis, we do not know the clustering information of the candidates, therefore we opted for unsupervised clustering.  

Unsupervised clustering is typically done by estimating the ``distance" or ``similarity" between different candidates, with the aim that candidates with low ``distance" might be similar and belong to the same cluster. 
The clustering algorithms we discuss in this paper use standard search pipeline features, without any expensive pre-processing, and can find reliable clusters in real-time. In some cases, physically meaningful relations between features could also be computed to enhance clustering performance, however we do not employ these here \citep[\S3.2]{pang2018}. 

The caveat to this approach is that unsupervised clustering techniques can be harder than supervised methods to tune for specific datasets. Also, due to the lack of true labels, it is difficult to evaluate the clustering performance. We discuss the above caveats further in Section~\ref{sec:methods}.

\subsection{Clustering RFI} 
Strong RFI events can also overwhelm the real-time pipelines by generating a large number of candidates, sometimes at all DM trials. Even though we use multiple filtration techniques to mitigate \hbox{RFI}, some signals still reach the pipeline's clustering step. Narrow-band RFI can lead to many candidates at all DMs in the DM grid (and, because of the resulting time shift of the peak, corresponding time bins). In some cases, the RFI appears as a strong localized source in the radio image and hence is present as a dense cluster of points in the image plane (for instance, this can happen with a sufficiently high-altitude satellite). Therefore, we can leverage the clustering algorithm to cluster those thousands of triggers into one cluster, reducing the computational load by orders of magnitude. As it is not feasible to manually label RFI examples into multiple clusters, we cannot evaluate clustering algorithms' performance on identifying separate RFI clusters. Instead, we only estimate clustering performance on FRB clusters. This is further discussed in Section~\ref{sec:metrics}.   



\section{Data}\label{sec:data}
Here we describe the details of our dataset and the features we used for clustering. We used \rf\ data to generate a dataset containing representative \hbox{RFI}, and used simulated FRBs to generate a dataset with representative candidates. We then combined the two datasets and applied four different pre-processing techniques (downsampling and normalization) on the features of the candidates to simulate 250 observations, consisting of candidates from both real and RFI events.  

\subsection{Feature Selection for Clustering}
\label{sec:features}
As mentioned in Section~\ref{sec:intro}, the pipeline reports a set of measured parameters for each candidate that satisfies the S/N threshold criterion.  
For our clustering analysis, we cluster based on DM, time (as is the standard with most past FRB searches), and sky position (with relative direction cosines represented by \emph{l} and \emph{m} as angular distances from the observation's pointing center) of the candidates. 

Candidates associated with an FRB event are expected to be densely located in \emph{l} and \emph{m}, as the FRB originates from a specific location in the sky. They would also show an expected S/N decrease in adjacent DM and time values (see Section~\ref{sec:whyclustering}) and would be closer for those parameters. On the other hand, RFI is randomly spread across this parameter space, but strong RFI can show a trend in DM and time if detected at multiple DMs.

\subsection{RFI database}
In this analysis, we used data from various commensal and commissioning observations of \rf\ system (project codes: 19A-242, 20A-330, 19B-223, 20A-163) in which the standard \rf\ pipeline detected only RFI candidates. This data spans a range of array configurations and other observing and search parameters (frequency, bandwidth, image pixel size, etc.). To create candidates representative of the real-time pipeline, we re-ran the \rf\ transient search on this data with the pipeline using default search parameters \citep[][]{law2018, law2020}.
As clustering performance is expected to be sensitive to the RFI environment, we selectively chose datasets with a variety of RFI types. 
These datasets are representative of the broad range of RFI we have seen at VLA and therefore form a robust sample of RFI for our analysis. This procedure was used to generate RFI candidates from 13 observations, with a few to $\sim6000$ candidates each. We manually verified that all these candidates were RFI and saved parameters relevant for clustering for each candidate (\S\ref{sec:features}). We will refer to this as the RFI dataset. 


\subsection{Simulating and Injecting FRBs}
\label{sec:sim_frbs}
We also generated a dataset of ``real'' candidates, representing our signals of interest. This was done by generating simulated data (with standard radiometer noise for different array configurations, observing, and search parameters) and injecting simulated transients. The distribution of parameters used for injecting transients is given in Table~\ref{tab:sim_props}. We searched this simulated data using \rf\ system with real-time search configuration to generate candidates. We saved the relevant parameters of all the candidates, and manually verified them to make sure that each candidate corresponds to the injected transient. We discarded any observation with less than four candidates. This procedure was used to generate real candidates from 114 simulated observations (with one transient injected in each observation). We will refer to this as the FRB dataset. 

\begin{deluxetable}{ccccc}
\tablecaption{Parameter distributions of simulated FRBs \label{tab:sim_props}}
\tablewidth{0.5\linewidth}
\tablehead{
\colhead{Parameter} & \colhead{Distribution} & \colhead{Range/Values}}
\startdata
S/N & Uniform  & 10, 40\\
DM (\dmunits) & Uniform & 10, max\_search\_dm\tablenotemark{a}\\
Width (ms) & Uniform &  1, 40 (ms)\\
Frequency & Uniform & L, S, C, X \\
Array configuration & Uniform & A, B, C, D\tablenotemark{b}\\
Sky position (l, m) & Uniform & -fov/2, fov/2\tablenotemark{c}\\
\enddata
\tablenotetext{a}{max\_search\_dm is the maximum DM searched for a given configuration
}
\tablenotetext{b}{Maximum baseline lengths for the four configurations (A, B, C, D) are 36.4\,km, 11.1\,km, 3.4\,km, 1.03\,km.}
\tablenotetext{c}{fov is the field of view at the randomly chosen frequency.}
\end{deluxetable}




\subsection{Test Dataset}
\label{sec:testdata}
To evaluate the performance of various clustering methods (described in Section~\ref{sec:algos}), we generated a test dataset. We used this dataset to compare the performance of different pre-processing techniques, and also during hyperparameter tuning (see Section~\ref{sec:hyperparam_tuning}). 

The test dataset consists of multiple observations, each containing some RFI candidates and some FRB candidates. We enforce that each observation has one transient event, and so all FRB candidates in an observation would be associated to that single event. Therefore, a perfect clustering algorithm should form only one FRB cluster per observation. To generate such an observation, we randomly chose one observation each from the RFI and FRB dataset pool. We then randomly select X\% of RFI candidates (where X is sampled from a uniform distribution between 20-100) from the RFI observation, all the FRB candidates from the FRB observation, and concatenate their features. We then randomise the order of the examples. This creates a set consisting of both RFI and FRB candidates. All observations with less than 10 total number of candidates were discarded. Using this process, we created 250 observations containing RFI and FRB candidates, which formed our test dataset.

\subsection{Pre-processing}
\label{sec:preprocessing}
Pre-processing is the procedure that takes the event features and converts them into indexed parameter ranges such that all the parameters will be equally weighted in terms of their importance in the clustering. 

As explained in Section~\ref{sec:features}, we use DM, time, \emph{l} and \emph{m} as features for clustering. Therefore, for each candidate in each observation of our database, we save these four parameters along with the image S/N of the candidate. After clustering, we use S/N to decide the representative candidate from each cluster. We convert the absolute value of DM to an index based on its index in the DM array for each candidate. Similarly, we also convert the time value (in seconds) to an index, based on the sample number corresponding to that time from the start of that processing segment. We also convert \emph{l} and \emph{m} (which is the offset of the candidate from primary beam center) to corresponding pixel values, using the synthesized beam size. Therefore, we convert all the features to corresponding indices. This is necessary as otherwise, different scales of different features might bias the distance estimates required in clustering.

The transient events we are interested in appear as a point source in the sky. Therefore, all the candidates from that transient should constitute a small range of \emph{l} and \emph{m} index values. We downsampled the \emph{l} and \emph{m} values of all the candidates to increase the sky density of candidates, which might enhance the clustering performance. We tried downsampling factors of 1, 2 and 4 (henceforth referred to as \texttt{DS1, DS2} and \texttt{DS4} respectively). Although we have scaled all the features to their corresponding indices, we also evaluated clustering performance on standardized data (i.e, with zero mean and unit variance, hereafter referred to as \texttt{Norm}). Throughout the paper, we report the performance of all the algorithms on all these different pre-processing techniques. We also try to determine the pre-processing technique which leads to the best clustering performance. 

\section{Methods}
\label{sec:methods}
\subsection{Clustering Algorithms}
\label{sec:algos}
We compare eight algorithms to cluster our test dataset: K-means, Mean Shift, Affinity Propagation, Agglomerative Clustering, DBSCAN, Optics, HDBSCAN, and Birch. For all except HDBSCAN, we use the implementation of these algorithms in \texttt{scikit-learn} \citep{scikit-learn, sklearn_api}. We use the python implementation of HDBSCAN by \citet[][]{mcinnes2017hdbscan}. We briefly discuss the details of these algorithms and their hyperparameters 
in Appendix~\ref{appendix:algos}. We refer the reader to the respective papers and \texttt{scikit-learn} documentation for more details.






\subsection{Hyperparameter Tuning}
\label{sec:hyperparam_tuning}
Each clustering algorithm has several input parameters that can be used to control the algorithm's clustering process and speed. These input parameters are called hyperparameters. Some algorithms are very sensitive to the choice of these hyperparameters while others are robust to a range of hyperparameters. Our aim is to find the hyperparameters for each algorithm, which leads to the best clustering performance (called optimal hyperparameters). The following three techniques are typically used to obtain the optimal hyperparameters: brute force grid search, random sampling, and Bayesian optimization. Grid search involves generating a grid of points covering the whole parameter space uniformly. The performance metric is then calculated on all the grid points, and the hyperparameter combination with the maximum value of the metric is chosen. In random sampling, the hyperparameter combinations are randomly chosen from a distribution of parameters. Bayesian optimization uses Bayesian techniques to parse the parameter space and obtain the optimal hyperparameter combination. 

Random sampling has been shown to be better than brute force grid search \citep{Bergstra2012}. This is because, in most cases, only a few hyper-parameters really matter, the importance of which changes with different datasets. This makes grid search a poor choice for searching for hyper-parameters for different datasets. Therefore, we opted to use random sampling. Also, because our parameter space is not very large, we decided not to use Bayesian optimization. 
Appendix~\ref{sec:hyperparameter_ranges} shows the ranges and various possible values of different hyperparameters that were tried for each algorithm. Whereever necessary, we used a random state of 1996 in the algorithms for reproducibility. 

\subsection{Performance Metric}
\label{sec:metrics}
The general idea of using a performance metric is to have a common reference point to rank the effectiveness of clustering algorithms (and their hyperparameters). The one with the maximum value of the metric has the best general performance. 



Critical to this idea is a clear statement of our goals. Our primary measurable goals with clustering are the following:
\begin{enumerate}
    \item {\bf Avoid missing a genuine event due to clustering}. This can happen due to over-aggressive clustering that identifies FRB candidates as false members of an RFI cluster. As only the highest S/N member from each cluster is processed further, 
    assigning FRB members to RFI clusters will lead to FRB candidate not passing further in the pipeline. This can happen if there are RFI candidates with S/N higher than that of the FRB candidates.    
    \item {\bf Each event of interest should be singly identified.} All candidates from one FRB event should be clustered into one cluster, rather than many small separate clusters representing a single event of interest. This is to minimize the number of candidates that are passed to the pipeline for post-processing and classification. 
\end{enumerate}
To represent these goals, we have developed the following metric. A higher value of the metric is favourable. We use \emph{homogeneity}, \emph{completeness}, \emph{v-measure} and \emph{recall} to calculate the metric (hereafter referred to as \emph{score}).  
In the following, an FRB cluster is defined as a cluster containing one or more FRB candidates, obtained after clustering. In the following discussion, we follow the terminology defined in Section~\ref{terminology}.

\subsubsection{Homogeneity}
Homogeneity is the measure of purity of the clusters with respect to true labels, i.e it estimates if each cluster contains only members of a single class (i.e either FRB or RFI). We calculate homogeneity for each observation in the test dataset. As we are primarily interested in performance on FRB candidates, and as RFI can be clustered into multiple clusters (for which we don't have true information), we define homogeneity only for FRB clusters in the observation. For each FRB cluster, we calculate the ratio of the number of FRB candidates in that cluster to total number of candidates in that cluster. Homogeneity (\emph{h}) is the weighted average of all these ratios, weighing them by the number of candidates in the cluster. Hence, 

\begin{equation}
\begin{aligned}
    h & = \frac{1}{N_T}\sum_i{\frac{n^i_{\!_{FRB}}}{n^i_T} n^i_T} \\
    &= \frac{1}{N_T}\sum_i{n^i_{\!_{FRB}}}
\end{aligned}
\end{equation}

where \emph{i} represents the $i$th FRB cluster, and the sum is over all the FRB clusters. $n^i_{\!_{FRB}}$ is the number of FRBs in the $i$th FRB cluster, $n^i_T$ is the total number of candidates in that cluster. $N_T = \sum_i n^i_T$ is the total number of candidates in all FRB clusters. \emph{h} can be between 0 (when all FRB candidates are left unclustered) and 1 (when all FRB clusters contain only FRB candidates).  

\subsubsection{Completeness}
Completeness is used to estimate if all members of a given class are assigned to the same cluster. We calculate completeness for each observation in the test dataset. We define completeness for FRB clusters, and a high completeness score will minimize the number of clusters the FRB candidates are clustered to. For each FRB cluster, we calculate the ratio of the number of FRB candidates in that cluster to the total number of FRB candidates. Completeness (\emph{c}) is equal to the weighted average of all these ratios, weighing them by the number of candidates in the cluster\footnote{We include unclustered candidates as a single cluster in this case, while unclustered candidates were ignored while calculating homogeneity.}. Hence, 

\begin{equation}
\begin{aligned}
    c & = \frac{1}{N_T}\sum_i{\frac{n^i_{\!_{FRB}}}{N_{\!_{FRB}}} n^i_T}
\end{aligned}
\end{equation}

where \emph{i} represents $i$th FRB cluster, and the sum is over all the FRB clusters. $n^i_{\!_{FRB}}$ is the number of FRBs in $i$th FRB cluster, $N_{\!_{FRB}}$ is the total number of FRB candidates in that observation. $N_T = \sum_i n^i_T$ is the total number of candidates in all FRB clusters. \emph{c} will be very small if all FRB candidates are assigned different clusters, and 1 when all FRB candidates are clustered into one cluster.

\subsubsection{V-measure}
V-measure is the harmonic mean between homogeneity and completeness. This is used as we want all the clusters to be pure and favor minimum number of clusters. Therefore, we want to maximize both homogeneity and completeness. V-measure (\emph{v}) will be 1 when both \emph{h} and \emph{c} are 1 and will be 0 if either of those is 0. Hence,  

\begin{equation}
\begin{aligned}
    v & = \frac{2 h c}{h + c }
\end{aligned}
\end{equation}

We calculate \emph{h, c, v} for each observation in the test dataset and take a weighted average of all \emph{v}'s (weighting by the total number of candidates in that observation) to get an estimate of V-measure for the whole dataset (\emph{V}).  

\subsubsection{Recall}
Recall is the fraction of FRBs that are recovered after clustering. 
After clustering, only the candidates with maximum S/N in each cluster are processed further in the pipeline. Therefore, if the clustering algorithm clusters FRB candidates together with high S/N RFI candidates, then the FRB will not be recovered from that cluster and might be missed. Therefore, recall (\emph{R}) is defined as the ratio of the number of observations in the dataset for which FRB was recovered 
to the total number of observations in the dataset. 

\subsubsection{Score}
Score is defined as the product of recall (\emph{R}) and total V-measure (\emph{V}). 

\begin{equation}
\begin{aligned}
    Score & = R \times V
\end{aligned}
\end{equation}

We use this score to compare the clustering performance of different algorithms and find the optimal hyperparameters for each clustering algorithm. 

\subsection{Advantages of this metric}
We have defined the above metric with respect to FRB and RFI clusters, but it can be easily generalized to any application with goals generally similar to those laid out in Section~\ref{sec:metrics}. This metric ensures that the information in relevant clusters is not missed by overaggressive clustering while still minimizing the number of clusters formed. There are some advantages of this metric over other clustering metrics available in the literature \citep[for a detailed comparison using a similar metric see][]{rosenberg-hirschberg-2007-v}: 1) It is independent of the clustering algorithm, size of the dataset, number of classes, and clusters. 2) It can appropriately use one (or more) base class (here FRB) to evaluate the clustering performance with respect to true labels, considering all data points of the relevant class. 3) Using homogeneity and completeness, V-measure gives importance to both pure clusters and minimum number of clusters. 4) By adequately weighting individual metrics, it is possible to concisely evaluate the clustering performance across multiple examples, in the form of a simple number (score).

\section{Results}\label{sec:results}
\subsection{Optimal Hyperparameters}\label{sec:hyperparams}


\begin{figure}
    \centering
    \includegraphics[width=0.5\textwidth]{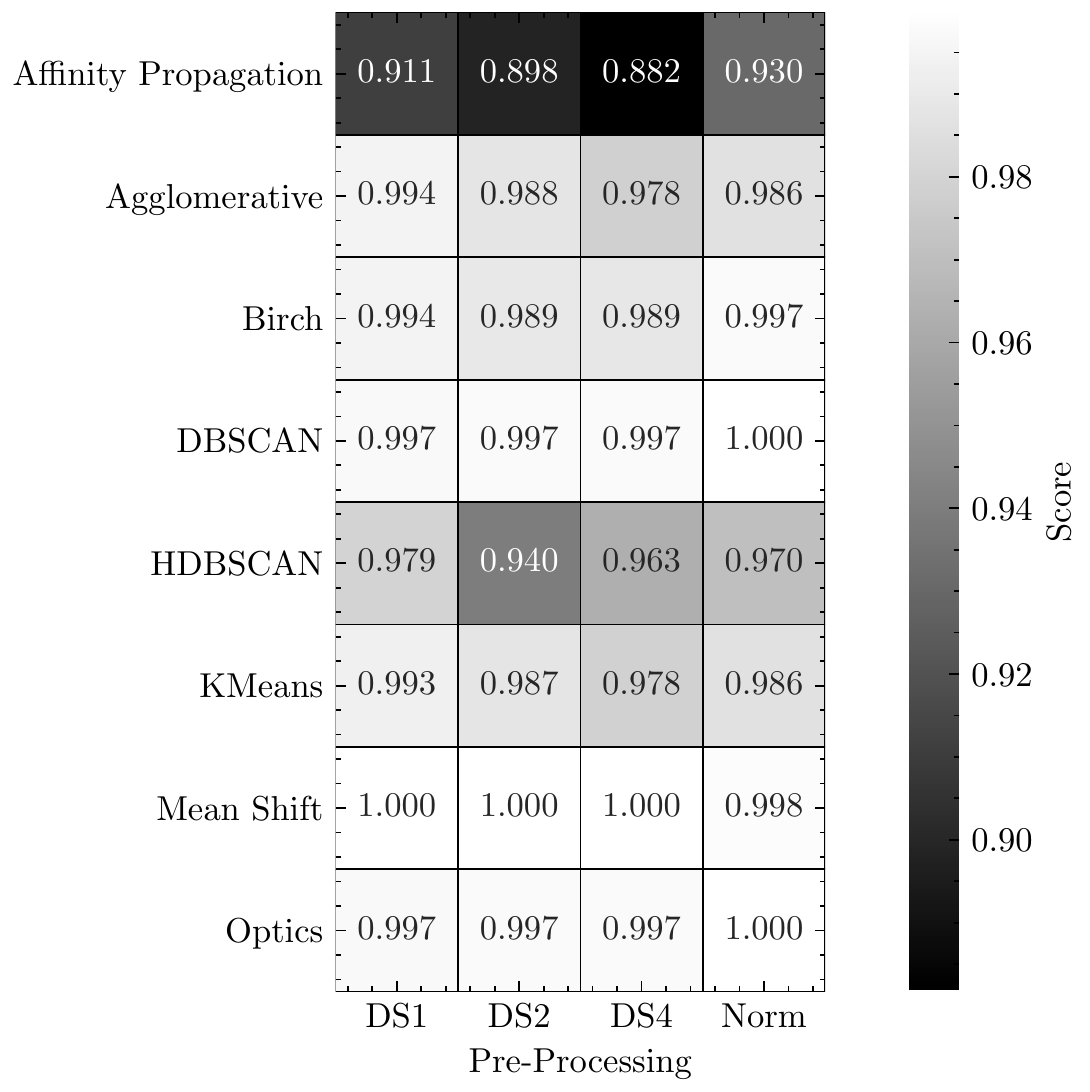}
    \caption{Maximum score obtained after random hyperparameter search (at the optimal hyperparameters) for each algorithm and pre-processing combination. Optimal hyperparameters for each case are given in Table~\ref{tab:besthyperparams}. DS refers to downsampling applied to \emph{l} and \emph{m} indices. Norm refers to normalization of the four features (see Section~\ref{sec:preprocessing})} 
    \label{fig:bestscorefig}
\end{figure}

We show the maximum score obtained for each algorithm after hyperparameter search in Figure~\ref{fig:bestscorefig}. Henceforth in this paper, we will refer to these hyperparameters as ``optimal hyperparameters". This figure shows that Mean Shift has the maximum score, out of all the eight algorithms. 
Table~\ref{tab:besthyperparams} shows these optimal hyperparameters for each algorithm.

\begin{deluxetable*}{cccccc}
\tablecaption{Optimal hyperparameters obtained for different algorithm and pre-processing combinations. \label{tab:besthyperparams}}
\tablewidth{\linewidth}
\tablehead{
\colhead{Algorithm} & \colhead{Hyperparameter} & \colhead{DS1} & {DS2} & {DS4} & {Norm}}
\startdata
Affinity Propagation  
& affinity & euclidean & euclidean & euclidean & euclidean  \\ 
& random\_state & 1996 & 1996 & 1996 & 1996  \\ 
& damping & 0.974 & 0.965 & 0.985 & 0.881  \\ 
& preference & -884 & -222 & -219 & -202  \\ 
\hline
Agglomerative  
& n\_clusters & 5 & 7 & 6 & 3  \\ 
& affinity & euclidean & manhattan & euclidean & euclidean  \\ 
& compute\_full\_tree & auto & auto & auto & auto  \\ 
& linkage & ward & average & ward & ward  \\ 
\hline
Birch  
& n\_clusters & 5.000 & 7.000 & 6.000 & 10.000  \\ 
& threshold & 0.341 & 0.876 & 0.676 & 0.957  \\ 
& branching\_factor & 13.000 & 56.000 & 85.000 & 77.000  \\ 
\hline
DBSCAN  & min\_samples & 2 & 2 & 2 & 2  \\ 
& eps & 14.163 & 14.726 & 14.615 & 1.082  \\ 
& metric & chebyshev & chebyshev & chebyshev & cityblock  \\ 
& algorithm & auto & auto & auto & auto  \\ 
& leaf\_size & 23 & 21 & 35 & 38  \\ 
\hline
HDBSCAN  & min\_samples & 5 & 5 & 5 & 5  \\ 
& metric & euclidean & euclidean & euclidean & cityblock  \\ 
& min\_cluster\_size & 2 & 3 & 2 & 9  \\ 
& cluster\_selection\_method & eom & eom & eom & eom  \\ 
& allow\_single\_cluster & True & True & False & True  \\ 
\hline
KMeans  & algorithm & full & elkan & full & auto  \\ 
& n\_clusters & 5 & 6 & 6 & 3  \\ 
& n\_init & 13 & 15 & 28 & 26  \\ 
& random\_state & 1996 & 1996 & 1996 & 1996  \\ 
\hline
Mean Shift  
& bandwidth & 16.416 & 32.750 & 19.350 & 1.229  \\ 
& bin\_seeding & True & False & True & True  \\ 
& cluster\_all & True & True & True & True  \\ 
\hline
Optics  & min\_samples & 2 & 2 & 2 & 2  \\ 
& eps & 14.782 & 14.376 & 14.551 & 1.095  \\ 
& metric & minkowski & chebyshev & minkowski & cityblock  \\ 
& min\_cluster\_size & 8 & 6 & 6 & 8  \\ 
& p & 14.672 & - & 11.009 & -  \\ 
& cluster\_method & dbscan & dbscan & dbscan & dbscan  \\ 
& xi & - & - & - & -  \\ 
\hline
\enddata
\end{deluxetable*}

\begin{figure*}
    \centering
    \includegraphics[width=\textwidth]{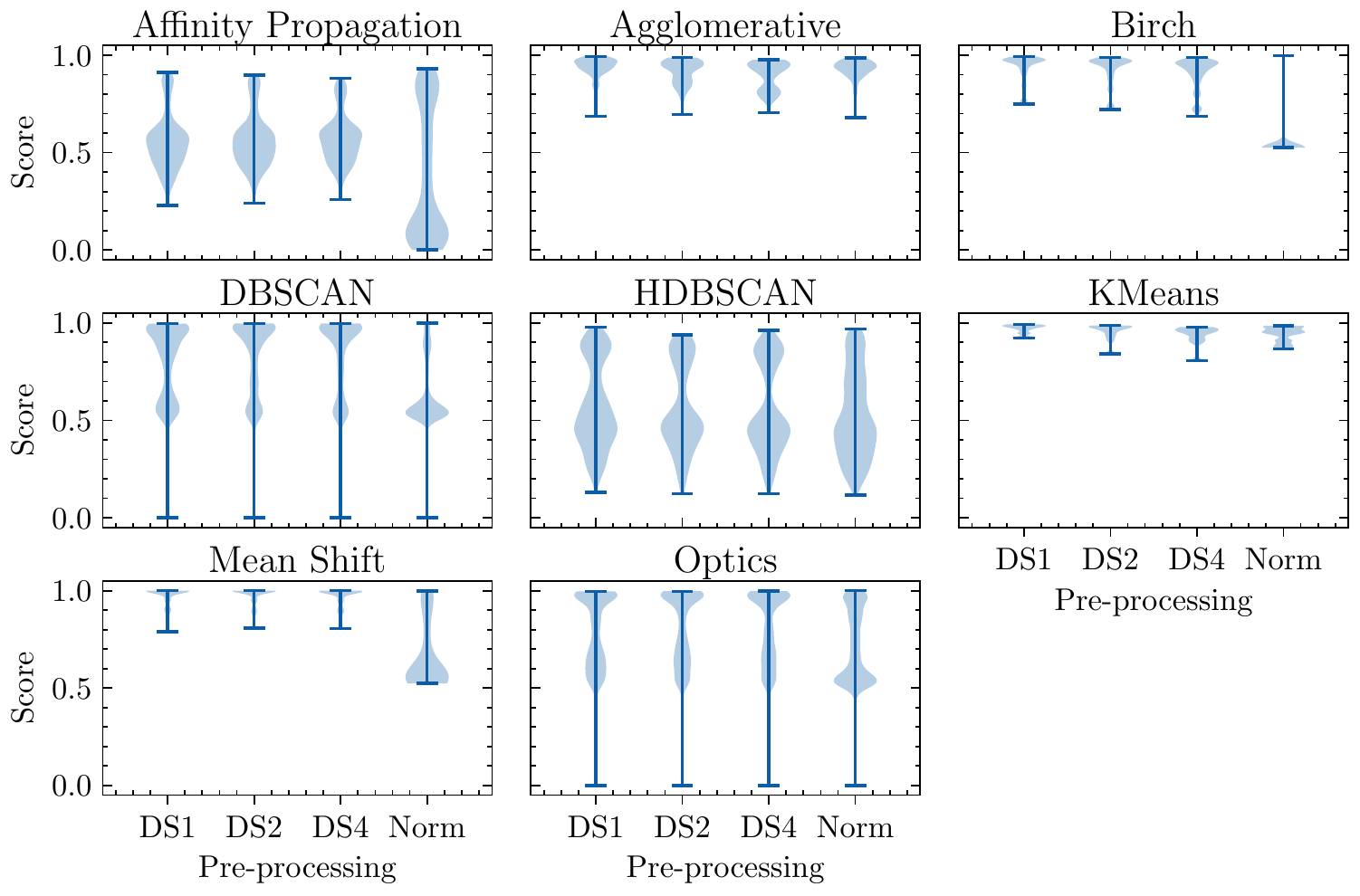}
    \caption{Violin plots of score vs. pre-processing for different clustering algorithms. Each violin plot shows the distribution of scores obtained at various hyperparameters evaluated during random hyperparameter search. Different sub-figures represent different algorithms (Sections~\ref{sec:hyperparam_tuning}, \ref{sec:hyperparams}). DS refers to downsampling applied to \emph{l} and \emph{m} indices. Norm refers to normalization of the four features (see Section~\ref{sec:preprocessing}).} 
    \label{fig:violin_grid}
\end{figure*}

Figure~\ref{fig:violin_grid} shows the distributions of scores at various hyperparameter values, for each combination of algorithm and pre-processing. 
Although not crucial to hyperparameter selection, the distribution of scores gives an insight into the robustness of the algorithm to the choice of input hyperparameters. In some cases, the scores vary between the full range of 0 and 1, while in others, the distribution is very narrow around high scores. The algorithms for which the score distributions peak around a high value should be more robust to the input hyperparameters than those for which the peak is at a middle or even low value of score. In the latter case, only a small range of hyperparameters would lead to a high score. 


\subsection{Effect of data processing}
As mentioned in Section~\ref{sec:preprocessing}, we also repeated the above experiment after pre-processing the data in two ways: downsampling the $l$ and $m$ indices and data normalization. We show the results for this in Figures~\ref{fig:bestscorefig} and \ref{fig:violin_grid}. As can be seen from Figure~\ref{fig:bestscorefig}, there is no clear trend of clustering performance for different pre-processing cases. We also note that the shape of the score distribution remains the same across different downsampling factors for a given algorithm. This indicates that downsampling doesn't have a significant contribution to the clustering performance. 

\subsection{Evaluating performance on clean data} 
\label{sec:cleandata}
So far, we have evaluated the performance of clustering algorithms on data with real RFI candidates along with simulated FRB candidates. This, as stated earlier, was a reasonable approximation of the candidates from real observations. 
Usually, in the case of a real transient, the pipeline only gets triggered at candidates from real transient, and no RFI is seen (either because of the amplitude of real transient or because low-level RFI is flagged). Therefore, here we report the performance of these clustering algorithms (at the optimal hyperparameters) on a dataset containing candidates only from a real event and no RFI. This is done to test the generalisability of these clustering algorithms on data without RFI. This would also serve as an independent test on unseen datasets for which the hyperparameters of the algorithms weren't tuned.

\subsubsection{Completeness on Clean data}
\begin{figure}
    \centering
    \includegraphics[width=0.5\textwidth]{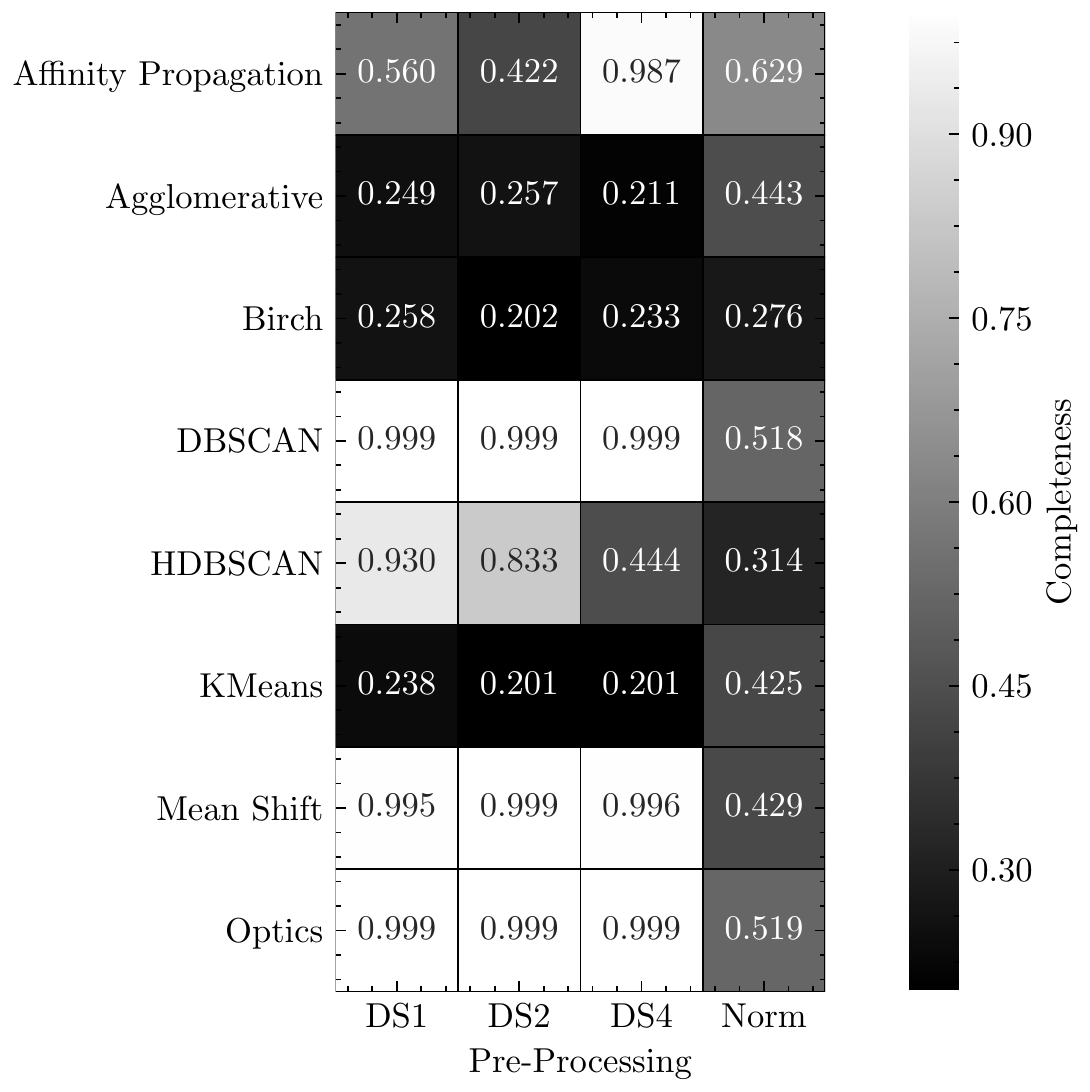}
    \caption{Completeness of different algorithms on clean data i.e without any RFI candidate (Section~\ref{sec:cleandata}). High completeness score is better and would imply that the FRB candidates are clustered in a minimum number of clusters for each of the 100 observations. Each algorithm was evaluated at its optimal hyperparameters (Table~\ref{tab:besthyperparams}). DS refers to downsampling applied to \emph{l} and \emph{m} indices. Norm refers to normalization of the four features (see Section~\ref{sec:preprocessing}). } 
    \label{fig:completeness_heatmap}
\end{figure} 

We use the same procedure as described in Section~\ref{sec:sim_frbs}, to generate a dataset of 100 observations with candidates from one simulated FRB each. We randomly chose the parameters of the simulated FRBs and observing configurations, as explained earlier and discarded any observation with less than ten candidates. 

We use \emph{completeness} (see Section~\ref{sec:metrics}) to report the clustering performance on this dataset. As there is no RFI candidate in this dataset, homogeneity would always be one and therefore is not a useful metric in this case. Here, a perfect clustering algorithm should generate just one cluster per observation for which completeness would be maximum, declining as the number of clusters increase. The overall completeness for a dataset is the average of all the completeness values from 100 observations, each weighted by the number of candidates in the observation.

Figure~\ref{fig:completeness_heatmap} shows the overall completeness score of each algorithm. DBSCAN, HDBSCAN, Mean Shift and Optics have the highest completeness score. It is to be noted that the completeness score of these four algorithms was worse when the data was pre-processed to zero mean and unit standard deviation (i.e \texttt{Norm}). On the contrary, downsampling the image features did not show any significant effect (with a notable exception of \texttt{DS4} for HDBSCAN).



\subsection{Benchmarking}
\label{sec:benchmarking}
\begin{figure}
    \centering
    \includegraphics[width=0.5\textwidth]{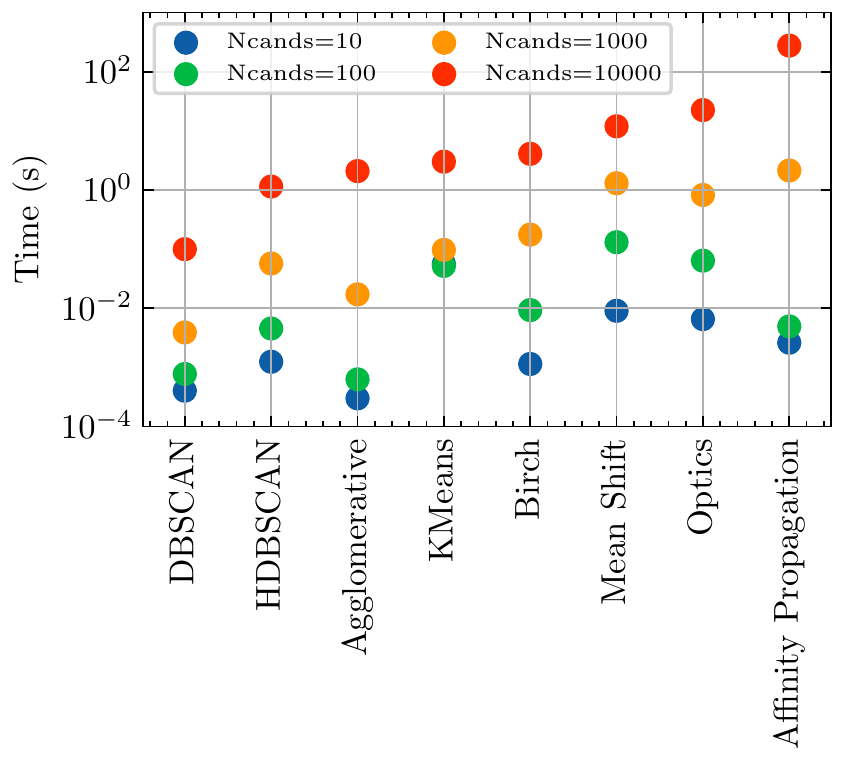}
    \caption{Time taken to cluster (in seconds) for each algorithm at their optimal hyperparameters. Different colors represent input data with different number of candidates. Results are shown only at optimal hyperparameters for \texttt{DS1}. DBSCAN and HDBSCAN are much faster than algorithms like Mean Shift and Affinity Propagation (Section~\ref{sec:benchmarking}).} 
    \label{fig:speed}
\end{figure}

We evaluated the clustering speed of all clustering algorithms at their optimal hyperparameters. To do this, we generated an observation with a varying number of candidates consisting of random values for the four features. We then ran all the clustering algorithms on those observations and recorded the time taken for just the clustering step. We did this test with optimal hyperparameters obtained for all four pre-processing cases. As the clustering speed is primarily dependent on the number of candidates to be clustered, we didn't use real data for this test. We show the result of this test in Figure~\ref{fig:speed}. The time taken did not vary significantly with parameters from different pre-processing cases, so we only show results using optimal hyperparameters for \texttt{DS1} in this figure. DBSCAN and HDBSCAN are the fastest of these algorithms, while Affinity Propagation, Mean Shift, and Optics are the slowest, by at least an order of magnitude.  

\section{Discussion}\label{sec:discussion}

\subsection{Feature Importance}
\label{sec:featureimp}
It is worth understanding the impact of feature selection on our outcome, as some features are expected to be more important than the others \citep[][]{featuresimp, featuresimp2}. 
We use a Random Forest classifier \citep[][]{randomforest}, implemented in \texttt{scikit-learn}, to estimate the relative feature importance of the four features, (DM, time, $l$, $m$), in determining accurate clusters.

\begin{figure}
    \centering
    \includegraphics[width=0.5\textwidth]{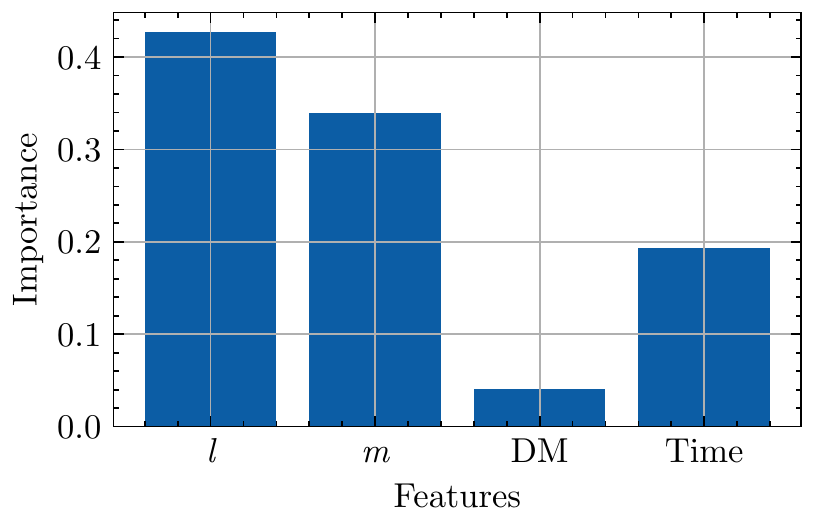}
    \caption{Importance of each feature, determined by training a Random Forest classifier to classify each observation into RFI and FRB. We trained the classifier individually on all observations in the test dataset and took a weighted average of the individual feature importance to obtain the above plot. \emph{l} and \emph{m} contribute much more towards classification than DM and Time (Section~\ref{sec:featureimp}).} 
    \label{fig:feature_importance}
\end{figure}

We used the test dataset (see Sec.~\ref{sec:testdata}) without any pre-processing, containing candidates from 250 observations (hereafter we refer to this data as \texttt{DS1}). We knew the true labels (RFI and FRB) for each candidate in those 250 observations. For each observation, we trained a Random Forest classifier (at the default input parameters) using all the candidates in that observation. From the trained classifier, we then used \texttt{feature\_importances\_} to obtain the relative feature importance of each of the four features. This attribute of the Random Forest classifier calculates Gini importance \citep[][]{randomforest} for each feature, which is representative of the importance of feature during classification.  We repeated this for all the observations in our test dataset. To estimate the total feature importance, we averaged all the importance for each feature weighing each by the number of candidates in that observation. The total importance obtained is shown in Figure~\ref{fig:feature_importance}. As can be seen from this figure, \emph{l} and \emph{m} (sky position indices of the candidate) contribute much more towards classification than the DM and Time indices of the candidates. 

A caveat to this simplistic analysis is that classification of all candidates into two classes, FRB and RFI, is not the same as clustering them into multiple clusters. The two cases would have been similar if all the RFI candidates could be assigned to a single cluster, which is not true. Therefore, even though this analysis shows that sky position contributes much more to classification, we suspect that the relative contribution of DM and Time for the clustering task would be higher than what is obtained here. 

\subsection{What if I only use DM and time for clustering?}
\label{sec:twofeature}
In \rf\ system, we search for transients on the radio image. Therefore, for each candidate we get DM, Time, \emph{l} and \emph{m} information. But in many experiments, typically the ones using a single dish telescope or the ones not performing an image-based transient search, only DM and time information is available for each detected candidate. Therefore, in those cases, only DM and time can be used to cluster the candidates together. 

We tested the clustering performance using only DM and time to cluster the observations in our test dataset. We used the optimal hyperparameters (listed in Table~\ref{tab:besthyperparams}) on the test dataset to evaluate this for all pre-processing cases. We also tested the clustering performance using only $l$ and $m$ indices to cluster our test dataset. As discussed in the previous section, the relative importance of sky positions is much higher than that of DM and time for a classification task. Therefore, clustering using only sky positions should give better scores than using just DM and time. 

\begin{figure}
    \centering
    \includegraphics[width=0.5\textwidth]{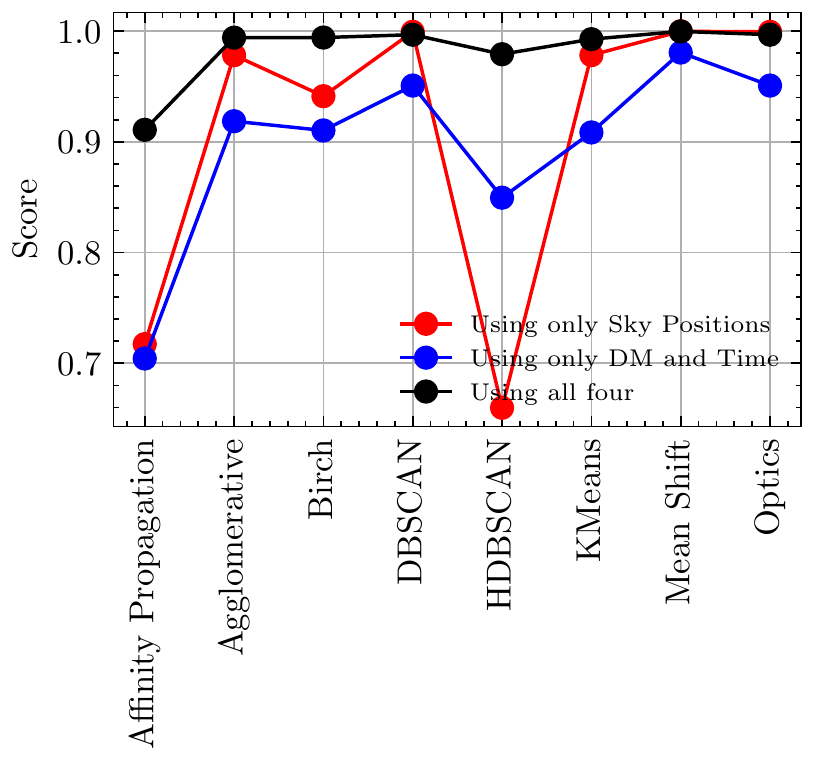}
    \caption{Score vs. algorithms for two feature clustering. Different colors represent different sets of features used to perform the clustering. We evaluated the scores on the test dataset. Results with \texttt{DS1} pre-processing are shown here (Section~\ref{sec:twofeature}).} 
    \label{fig:two_feature_clustering}
\end{figure}

We show the results of these two tests in Figure~\ref{fig:two_feature_clustering} along with the scores when all four features are used for clustering. We only show scores for one pre-processing case (\texttt{DS1}), as results with other pre-processing techniques were also similar. As can be seen from this figure, scores obtained using just sky positions (red curve) or DM and Time (blue curve) follow each other closely. Using sky positions shows minor improvement in score for most of the algorithms. Using all four features, as expected, gives the highest score that is $\sim$\,10\% better than the other two cases. 

This test highlights the importance of using sky positions along with the standard DM-time features to identify clusters of candidates originating from the same event. Therefore, if the sky position information is available for a candidate, it should also be used while clustering in the pipeline. With more and more interferometers (like ASKAP and DSA-110) implementing a \rf\ like search for transients on radio images in the future, it would be useful for them to incorporate sky position information to cluster candidates in their respective pipelines.

As careful readers would have noticed, a caveat to this test is that in clustering with two parameters, we didn't do a hyperparameter search to obtain the optimal hyperparameters that maximize the score using those two features. Instead, we used the hyperparameters that were optimal when four features were used. A full hyperparameter search using two parameters might lead to a different set of parameters that might improve the score further. But even with this simple test, it can be noted that only using sky positions for clustering gives an improvement in score in almost all cases. 

We have demonstrated in this and the previous section that sky positions are overall more important for clustering than DM and Time. This could be because RFI candidates are more likely to span a wide range of time and DM values, which might overlap with those of FRB candidates, while they are still localized in the radio image. 
Therefore, it is less likely (though still possible) for RFI to be very close to an FRB in the radio image. Similarly, RFI may be highly variable in frequency/time space, whereas in a radio image even unfocused (near-field) RFI will show up as contiguous streaks or other similarly structured patterns in images. Regardless of the reason for this, however,, we have demonstrated here that when possible, sky positions should be used for clustering candidates.

\subsection{But which algorithm should I use?}
There are several considerations when deciding what algorithm to use based on the comparative analysis we have presented here.

\begin{itemize}
    \item \textbf{Maximum Score}: As discussed in Section~\ref{sec:metrics}, we want the clustering algorithm to meet our application-specific goals; of not missing a genuine event and singly identifying FRB candidates. Our performance metric (called score) maximizes when these goals are met. Therefore, we could search for a set of optimal hyperparameters for each clustering algorithm, that gives the maximum score. All algorithms, except Affinity Propagation, have an optimal score above 0.95 (Figure~\ref{fig:bestscorefig}). 
    
    \item \textbf{Generalisable}: The clustering algorithm needs to generalize to various types of data it can encounter in the pipeline. By testing the algorithms and optimal hyperparameters obtained in the previous step on an independent dataset, one could quantify the algorithms' generalisability. To be more application-specific, we tested this on a dataset with observations containing candidates only from a real event, without any RFI, and computed the completeness as the performance metric. Only four algorithms, DBSCAN, HDBSCAN, Mean Shift, and Optics gave completeness above 0.9 in this test (Figure~\ref{fig:completeness_heatmap}). 
    
    \item \textbf{Speed}: Finally, the clustering algorithm would only have a limited amount of time to cluster candidates. Therefore, even for a large number of candidates, it should not exceed the limited time constraint. In our specific application for \rf, clustering is performed on candidates generated from small segments of data that are tens of seconds long. Based on the other pipeline steps, clustering shouldn't take longer than a few seconds. The number of candidates detected by the search step typically varies between a few to thousands of candidates for a segment. Based on these requirements DBSCAN, HDBSCAN, Agglomerative, and K-means can be used (Figure~\ref{fig:speed}).  
\end{itemize}

As an example using the \rf\ system, selecting the algorithms using the above three steps, we conclude that either DBSCAN or HDBSCAN can be used for clustering \rf\ data. Based on the results in Figures~\ref{fig:completeness_heatmap} and \ref{fig:two_feature_clustering} we can further infer that DBSCAN is better than HDBSCAN. As reported earlier, we didn't notice any improvement by using different pre-processing techniques, therefore no pre-processing is favored (Figure~\ref{fig:bestscorefig}). A similar procedure can also be used to choose the clustering algorithm for any other single-pulse search pipeline or even for a more general clustering application.

\section{Conclusions}\label{sec:conclusions}
In this paper, we have compared eight different unsupervised algorithms to cluster candidates generated by single-pulse search pipelines. We have also analyzed the effects of various pre-processing techniques on the data. We used real RFI from \rf\ system and simulated FRB candidates to test different algorithms. We have developed a performance metric to quantify clustering performance. This metric makes sure that FRBs are not missed due to overaggressive clustering while still minimizing the number of clusters formed. Using a random hyperparameter search, we obtained optimal hyperparameters, which maximizes this metric for different algorithms. We test all the algorithms with optimal hyperparameters on an independent dataset consisting of only FRB candidates to evaluate the generalisability of different algorithms. We also estimated the average clustering time for various algorithms on a dataset of varying sizes. Finally, we have proposed a strategy that can be used to choose a clustering algorithm, using various tests mentioned earlier. We apply this strategy to obtain a clustering algorithm appropriate for \rf\ system. This strategy can also be used at other single-pulse search systems to obtain the optimal clustering algorithm. Our strategy is generic enough to be used for other clustering applications. Our performance metric can also be used in other clustering applications where clustering information for only one cluster of interest is available, out of an unknown number of true clusters. We have also demonstrated that using spatial features for clustering improves the clustering performance compared to the traditional approach of just using DM and time features. All the scripts used in this analysis are openly available in a Github repository\footnote{\url{https://github.com/KshitijAggarwal/rfclustering}}.

\section*{Acknowledgements}
K.A. would like to thank Shalabh Singh for useful discussions regarding the performance metric. K.A. and S.B.S acknowledge support from NSF grant AAG-1714897. SBS is a CIFAR Azrieli Global Scholar in the Gravity and the Extreme Universe program.
Part of this research was carried out at the Jet Propulsion Laboratory, California Institute of Technology, under a contract with the National Aeronautics and Space Administration.
The NANOGrav project receives support from National Science Foundation (NSF) Physics Frontiers Center award number 1430284.
The National Radio Astronomy Observatory is a facility of the National Science Foundation operated under cooperative agreement by Associated Universities, Inc. 
 
\facilities{EVLA}

\software{numpy \citep{numpy}, matplotlib \citep{Hunter:2007}, pandas \citep{pandas, scipy}, scikit-learn \citep{scikit-learn, sklearn_api}, hdbscan \citep{hdbscan}, rfpipe \citep{rfpipe}}

\bibliography{rfcluster} 

\appendix
\section{Clustering Algorithms}
Here we give a brief overview of all the clustering algorithms used in this analysis and some details and potential advantages/disadvantages of each algorithm for our clustering application. 

\label{appendix:algos}
\subsection{K-means}
K-means \citep{kmeans} algorithm is one of the most widely used clustering algorithms. Given an input number of clusters, the algorithm randomly initializes centroids for each cluster. Each example is then assigned a cluster based on the distance from that centroid. A new centroid is then computed for each cluster, and all examples are reassigned to the new centroids. 
This process is repeated till a convergence criterion is met. The main challenges with K-means are that it is not good at identifying non-spherical clusters and requires the number of clusters as input, both of which limit its ability to generalize on different datasets. 


\subsection{Mean Shift}
Mean Shift \citep{meanshift} is a centroid based algorithm.
The algorithm assumes that the data is drawn from an underlying probability density function and tries to estimate it using Kernel Density Estimation. Then, it calculates a centroid for each data point using the kernel and iteratively updates the centroid using a mean shift-vector. 
At convergence, the centroid will be placed at the nearest highest density peak of the density function. The same process is repeated for each data point, and the data points which lead to the same high-density peaks are then assigned to the same cluster. 
The only hyperparameter here is the bandwidth of the kernel. 
Mean Shift is not highly scalable as it requires multiple nearest neighbor searches. 


\subsection{Affinity Propagation}
Affinity Propagation \citep{affinity_propagation} is based on the concept of ``message passing" between data points. It tries to find ``exemplars", i.e, members that are representative of clusters. 

It starts by calculating a similarity matrix, which can be defined as the negative squared distance between two data points. The diagonal of this matrix is set to a constant, called ``preference", which is an input hyperparameter. Preference determines how likely a particular data point would be to become an exemplar. The algorithm then calculates three matrices, called Responsibility, Availability, and Criterion Matrix. 
These matrices are updated iteratively till a convergence criterion is met, and then clusters are assigned based on the information in Criterion Matrix. The details of the algorithm are given in \citet{affinity_propagation}. 
Affinity Propagation's advantage is that it doesn't require the number of clusters as input, but the algorithm is computationally complex and can be slow on large datasets. 







\subsection{Agglomerative clustering}
Agglomerative clustering \citep{agglo} is a type of hierarchical clustering. Hierarchical clustering algorithms start with each example being a different cluster and then merge the ones that are closer until there is only one cluster. Therefore, they can form a hierarchy of clusters (at various distances), which is represented as a tree. A linkage criterion \citep[see Section 5.1 of][]{jain1999} is used to decide the merge strategy. 
To determine the clusters from this cluster hierarchy, one has to choose a level or a cut in the tree. 
As was the case with K-means, the main challenge with this algorithm is to choose the number of clusters. 


\subsection{DBSCAN}
Density Based Spatial Clustering of Applications with Noise, or DBSCAN \citep{dbscan}, is a density-based clustering algorithm. It assumes that clusters lie in dense regions. It primarily requires two input hyperparameters: a density threshold (\textit{MinPts}) of a core point and a radius ($\epsilon$) of its neighborhood. A point that has at least \textit{MinPts} adjacent points in its $\epsilon$ neighborhood is considered a core point. Core points and their neighborhood are considered dense regions that form clusters, and overlapping dense regions are merged into a single cluster. Any point that is neither a core point nor falls within the neighborhood of a core point is classified as noise. It doesn't require the number of clusters as input, although the clustering output is very sensitive to other input parameters.  


\subsection{Optics}
Ordering Points To Identify the Clustering Structure or Optics \citep[][]{optics}, is a density-based clustering algorithm. Similar to DBSCAN, Optics requires two hyperparameters: $\epsilon$ and \textit{MinPts}, although  $\epsilon$  is not necessary. 
It uses the following distances: core distance (minimum radius required to classify a given point as core point) and reachability distance (calculated by comparing the distance between two core points and their core distances) to order points. The reachability distance for points in a cluster would be low. The Optics algorithm builds a reachability graph, which assigns each sample a reachability distance. 
A post-processing procedure is applied to the reachability plot to determine clusters. This procedure can be very sensitive to the input parameters. 
An advantage of Optics is that it can find clusters of varying density. Like other density-based algorithms, Optics doesn't require the number of clusters as input and can also find non-spherical clusters.   



\subsection{HDBSCAN}
Hierarchical Density Based Spatial Clustering of Applications with Noise or HDBSCAN \citep[][]{hdbscan, mcinnes2017hdbscan, mcinnes2017accelerated}, is very similar to Optics, i.e., it takes the approach of DBSCAN but extends it by varying the values of $\epsilon$. 
It forms a hierarchical tree that shows the clustering output. By parsing through the tree, going from one large cluster to many smaller clusters, HDBSCAN constructs a tree with persistent clusters based on its only hyperparameter: minimum cluster size. It then uses a stability criterion to extract the final clusters from the cluster tree. Like Optics, HDBSCAN can also form clusters of varying density and do not require the number of clusters as input.     


\subsection{Birch}
Balanced Iterative Reducing and Clustering using Hierarchies or Birch \citep[][]{birch} is a hierarchical clustering algorithm used typically on very large datasets. It is local, which means that the clustering decision is made without scanning all data points and existing clusters. 
It uses a Clustering Feature (or CF) which consists of summary of statistics for a given sub-cluster. 
CF is used to calculate the distance between two sub-clusters.
It creates a CF Tree consisting of these CFs. The Birch algorithm has two hyperparameters: branching factor and threshold; the former limits the number of CFs in a node of CFT, while the latter limits the distance for a new sample to be a part of an existing CF. The terminal nodes of a CFT are then clustered using another clustering algorithm to obtain final clusters. 






\section{Parameter Ranges for Hyperparameter tuning}
\label{sec:hyperparameter_ranges}
\label{tab:hyperparams}
\begin{deluxetable}{ccc}
\tablecaption{Hyperparameter ranges explored for different clustering algorithms using random sampling. Random uniform sampling was used to sample hyperparameters for all the parameter ranges/values.}
\tablewidth{\linewidth}
\tablehead{
\colhead{Algorithm} & \colhead{Hyperparameter} & \colhead{Range/Values}}
\startdata
Affinity Propagation & affinity & euclidean \\
& random\_state & 1996 \\ 
& damping & 0.5, 1 \\
& preference & -1000, -200 \\
\hline 
Agglomerative & n\_clusters & 2, 10 \\
& affinity & euclidean, manhattan, cosine \\
& compute\_full\_tree & auto \\
& linkage & complete, average, single, ward\tablenotemark{a} \\ 
\hline
Birch & n\_clusters & 2, 10 \\
& threshold & 0.1, 20 \\
& branching\_factor & 10, 100 \\ 
\hline
DBSCAN & min\_samples & 2, 10\\
& eps & 0.5, 15\\
& metric & euclidean, chebyshev, cityblock,\\
& & manhattan, canberra, hamming\tablenotemark{b} \\
& algorithm & auto \\
& leaf\_size & 20, 40\\
\hline
HDBSCAN & min\_samples & 2, 5 \\
& metric & euclidean, chebyshev, cityblock,\\
& & manhattan, canberra, hamming \\
& min\_cluster\_size & 2, 10  \\
& cluster\_selection\_method & eom, leaf \\ 
& allow\_single\_cluster & True, False \\
\hline
KMeans & algorithm & auto, full, elkan \\
& n\_clusters & 2, 10 \\
 & n\_init & 10, 30 \\
 & random\_state & 1996 \\
\hline
Mean Shift & bandwidth & 10, 40\tablenotemark{c} \\
& bin\_seeding & True, False \\
& cluster\_all & True, False \\
\hline 
Optics & min\_samples & 2, 10\\
& eps & 0.5, 15\\
& metric & minkowski, euclidean, chebyshev, canberra, \\
& & cityblock, manhattan, hamming\tablenotemark{b}\\
& min\_cluster\_size & 2, 10\\
& p & 1, 15\tablenotemark{d}\\
& cluster\_method & dbscan, xi\tablenotemark{e}\\
& xi & 0, 1\\
\enddata
\tablenotetext{a}{ward only works with euclidean affinity}
\tablenotetext{b}{eps range of 0.1 to 1 was used with hamming metric, and a range of 0.1 to 4 was used with canberra metric}
\tablenotetext{c}{bandwidth of 1 to 10 was used for normalised pre-processing case}
\tablenotetext{d}{p was used only with minkowski metric}
\tablenotetext{e}{Value of eps was used only with dbscan method, and value of xi was used when xi was selected as cluster method}
\end{deluxetable}

\end{document}